\documentclass[floatfix, prl, twocolumn]{revtex4}
\usepackage{graphicx,hyperref}

\newcommand\Tc{$T_\mathrm{c}$}

\begin{document}
\title{Andreev reflection at high magnetic fields: Evidence for electron and hole transport in edge states}
 \date{\today}
\author{J. Eroms}
\altaffiliation[Present address: ]{Kavli Institute of Nanoscience, Delft University of Technology, 
Lorentzweg 1, 2628 CJ Delft, The Netherlands}
\email{eroms@qt.tn.tudelft.nl}
\author{D. Weiss}
\affiliation{Universit\"at Regensburg, Institut f\"ur Experimentelle und Angewandte Physik, D-93040 Regensburg, Germany}
\author{J. De Boeck}
\author{G. Borghs}
\affiliation{IMEC, Kapeldreef 75, B-3001 Leuven, Belgium}
\author{U. Z\"ulicke}
\affiliation{Institute of Fundamental Sciences and MacDiarmid Institute for Advanced
Materials and Nanotechnology, Massey University,
Private Bag 11~222, Palmerston North, New Zealand}
\begin{abstract}
We have studied magnetotransport in arrays of niobium filled grooves 
in an InAs/AlGaSb heterostructure. The critical field of up to 2.6 T
permits to enter the quantum Hall regime. In the superconducting state, we observe strong magnetoresistance oscillations,
whose amplitude exceeds the Shubnikov-de Haas oscillations by a factor of about two, when normalized to the background. Additionally, we find that 
above a geometry-dependent magnetic field value the sample in the superconducting state has a higher longitudinal 
resistance than in the normal state. Both observations can be explained with edge channels populated with 
electrons and Andreev reflected holes. 
\end{abstract}
\maketitle

% Introduction

The analysis of superconductor-semiconductor structures has been an 
active field of research in recent years (see, e.g.,
Ref.~\onlinecite{Bib:SchaepersHabil} and references therein). 
The versatility of 
semiconductors and the high mobilities attainable in heterostructures in 
combination with the retroreflecting and phase coherent process of Andreev 
reflection~\cite{Bib:Andreev1} have allowed to observe a number of unique phenomena.
By now, 
experiments in the regime of low magnetic fields, i.e. no larger than a 
few flux quanta per junction area, are well established. 
Gateable Josephson currents~\cite{Bib:JoFET}, 
quasi-particle interference~\cite{Bib:Bastian}, 
phase coherent oscillations~\cite{Bib:Morpurgo1} 
and an induced superconducting gap~\cite{Bib:Merkt1} have been 
observed, to name a few. 

In the high-field regime, experimental evidence is much less abundant. 
A number of theoretical papers have dealt with Andreev reflection at high 
fields~\cite{Bib:Zuelicke,Bib:Asano,Bib:HochB,Bib:Takagaki}. 
Notably, Ref.~\onlinecite{Bib:Zuelicke} describes how edge channels
in the quantum Hall regime are formed of electron and hole states. To 
enter the regime of a fully developed quantum Hall effect external fields 
of several Tesla are required. Experiments have been performed with high 
critical field superconductors, such as NbN~\cite{Bib:NbN,Bib:Batov} or sintered 
SnAu~\cite{Bib:Moore}, each of 
which suffer from technological difficulties, making the interpretation 
of the experiments in the quantum Hall regime difficult. In this work, 
however, we report clear evidence of the influence of Andreev reflection 
on transport in edge states using the well established Nb-InAs system.
The critical field of up to 2.6~T permits to enter the quantum Hall
regime at high filling factors.

For the sample geometry we have chosen an array of niobium filled grooves 
in an InAs-AlGaSb heterostructure containing a high-mobility 
two-dimensional electron gas (2DEG). A similar arrangement has been 
studied previously~\cite{Bib:Drexler,Bib:Correa} in low magnetic 
fields. An important difference to single S-2DEG-S
junctions is that the voltage probes are located in the 2DEG. 
\begin{figure}
\includegraphics[width=8.1cm]{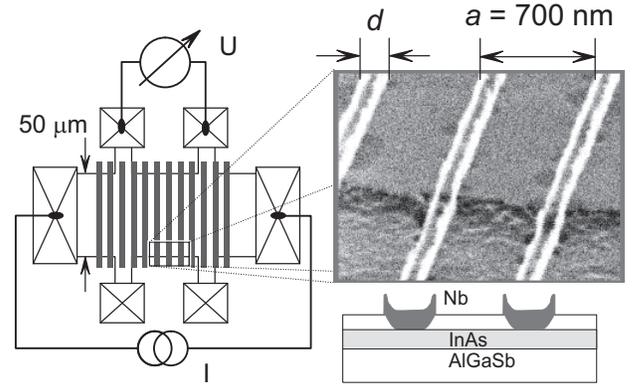}
\caption{Left: Geometry for the four-point measurements. Right: A
scanning electron micrograph of the sample taken at the mesa edge. 
A cross section of the sample is also shown.}
\label{Fig:SEM}
\end{figure}

% Sample fabrication

The samples were fabricated from a high-mobility InAs/AlGaSb quantum well,
which was grown by molecular beam epitaxy on a GaAs substrate~\cite{Bib:Behet}. 
Mesas of 50 $\mu$m width and Ohmic contacts were
prepared using optical lithgraphy. After this step an electron density of $n_s = 1.25\times 10^{12}$\ cm$^{-2}$
 and a mobility of $\mu = 200000$\ cm$^2$/Vs were found. 
The mean
free path in this material was therefore 3.8~$\mu$m, allowing for
ballistic transport in nanostructures. The Nb-filled grooves were  
defined with electron beam lithography, selective reactive ion etching 
(RIE) of the top AlGaSb layer, niobium sputter deposition and lift-off.
An {\em in situ} argon ion etch prior to the Nb sputtering ensured a 
high transparency of the Nb-InAs interface ($Z=0.63$ in the
OTBK-model~\cite{Bib:OTBK}), which allowed to observe several subharmonic 
gap structures in the differential resistance at low magnetic fields.
More fabrication details can be found in~\cite{Bib:Eroms}. 
A scanning electron
micrograph and a schematic cross section of the sample are shown in
Fig.~\ref{Fig:SEM}. Lattice periods were ranging from $a=400$~nm to $a=3\
\mu$m with different Nb-stripe widths. 

% Measurement setup and magnetotransport

\begin{figure}
\includegraphics[width=8.6cm]{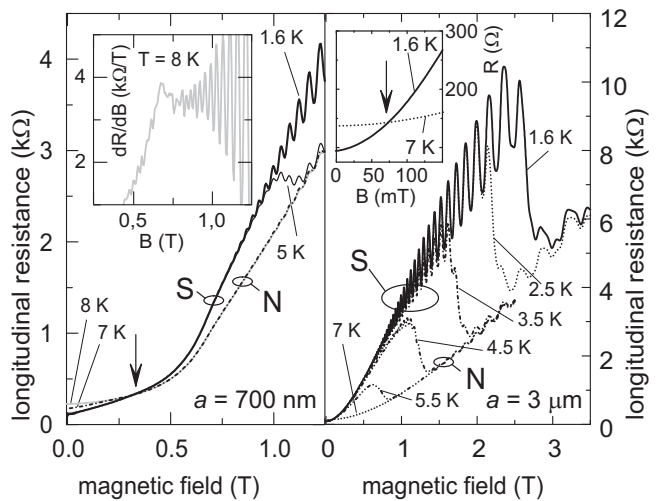}
\caption{$T$-dependent magnetotransport curves for two different samples. 
 Arrows: crossing points of the
graphs above \Tc\ and below \Tc . Left inset: The peak in $dR/dB$ corresponds to a
change in slope of the resistance trace. Right inset: Enlarged view of the
crossing point for the sample with $a = 3\ \mu$m. The critical temperatures were 7.4~K (left) and 6.9~K (right). Letters `N' and `S' denote the normal and superconducting branch, respectively.}
\label{Fig:TCurves}
\end{figure}

The magnetoresistance measurements were done in a four-point configuration, but
given the periodic geometry of the sample we effectively measured a series connection of
many S-2DEG-S junctions. The critical temperature \Tc\ of the Nb stripes was ranging from
6.9~K to 8.3~K, depending mainly on the stripe width. Figure~\ref{Fig:TCurves} shows
the magnetotransport curves of two samples with lattice periods $a=700$~nm and
$a=3\ \mu$m. The Nb stripes were 120~nm wide and 70~nm thick in both cases.

 Except for very low fields where the proximity effect dominates, the
curves lie on one of two branches, depending on whether the niobium stripes
are in the normal or superconducting state. A
transition between both branches is observed in 
Fig.~\ref{Fig:TCurves} when the critical field of
the niobium stripes at a given temperature is surpassed. Both
branches cross at a certain magnetic field (arrows in 
Fig.~\ref{Fig:TCurves}). At low fields,
the resistance on the superconducting branch is lower than on the normal
branch, as expected for high quality contacts. For high fields however, the
magnetoresistance in the superconducting state is higher than in the normal state.
This behaviour is not due to a low contact transparency, 
resulting in a reduced Andreev reflection 
probability. In that case 
the resistance below \Tc\ would {\em always} be higher than above \Tc . A crossing point would not be observed. 
In a reference sample where the contact transparency was 
deliberately reduced, the resistance in the superconducting
state was indeed higher than in the normal state throughout the 
entire field range. 

\begin{figure}
\includegraphics[width=7.5cm]{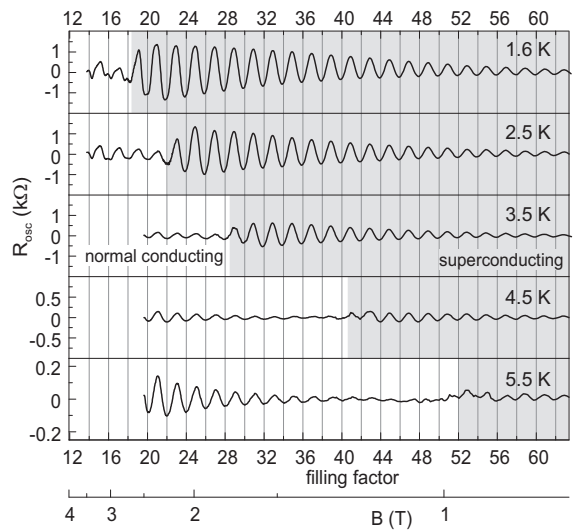}
\caption{Same data as in Fig.~\ref{Fig:TCurves}, right, after subtracting the slowly varying background and plotted against the filling factor. For ease of comparison, the value of $B$ is also given. Shaded regions: Nb stripes are superconducting, as extracted from Fig.~\ref{Fig:TCurves}. Note the strong increase of the amplitude at the superconducting transition.}
\label{Fig:invB}
\end{figure}

At $B>1$~Tesla, oscillations appear in the magnetoresistance. 
In a two-dimensional electron system, magnetotransport 
oscillations are observed as soon as the magnetic field is 
strong enough such that Landau level quantization is resolved 
experimentally.
Since the critical field of the niobium lines is much higher than the onset of the oscillations, 
we observe the impact of 
Andreev reflection 
on transport in the quantum Hall regime. On the superconducting branch, the
oscillation amplitude is much more pronounced than on the normal branch. This can be seen more clearly in Fig.~\ref{Fig:invB}, 
where the data from Fig.~\ref{Fig:TCurves}, right has been replotted versus $1/B$, after subtracting the slowly varying 
part of the magnetoresistance. The increase in amplitude is indeed quite striking. The two main experimental findings in our samples are therefore the higher resistance at high fields and the strong 
increase in the amplitude of the $1/B$-periodic oscillations.

%Magnetoresistance oscillations
Let us first consider the magnetoresistance oscillations in more detail. We evaluated the increase in amplitude for samples with a lattice period $a$ of 1\ $\mu$m, 2\ $\mu$m, and 3\ $\mu$m (data of the latter sample are shown in Fig.~\ref{Fig:TCurves}, right). The oscillation amplitude in the superconducting case was higher by a factor of 1.45, 4.1, and 6.4, respectively, when the amplitudes slightly above and below $B_\mathrm{c}$ were compared at $T=3$~K. Thus, the further the stripes were apart, the more striking the increase in amplitude. The higher oscillation amplitude is not simply due to the larger non-oscillatory resistance in the superconducting state, caused e.g. by the higher conductivity of the Nb stripes in the superconducting state. Normalized to the increase of the background, the oscillation amplitude still increased by 
a factor of up to 1.9 in the superconducting case~\cite{Note:Normalize} and the dependence on the stripe 
separation was maintained.

%When normalizing the increase in amplitude to the relative increase of the magnetoresistance, we still found that the 
%oscillations were stronger by a factor of up to 1.9 in the superconducting case and the dependence on the stripe 
%separation was maintained.

We also fitted the temperature dependence of the oscillation amplitude~\cite{Bib:Ando}. In the normal conducting case the amplitude was well described by thermal activation over the Landau gap and gave an effective mass of $m_\mathrm{eff}=0.04\ m_0$, where $m_0$ is the free electron mass. The same effective mass was found in a sample with the same geometry, but Au stripes instead of niobium. This value is comparable to what is found in InAs-based 2DEGs with a high carrier density~\cite{Bib:meff}. In the superconducting case however, the fit was poor and yielded an effective mass of up to $0.1\ m_0$, which is far from the real value. Therefore, Landau level splitting alone cannot be the underlying mechanism (and the effective mass extracted from such a fit is meaningless). Instead, edge channels containing electrons and Andreev-reflected holes can lead to the enhanced oscillations, as we discuss now. 

\begin{figure}
\includegraphics[width=8.6cm]{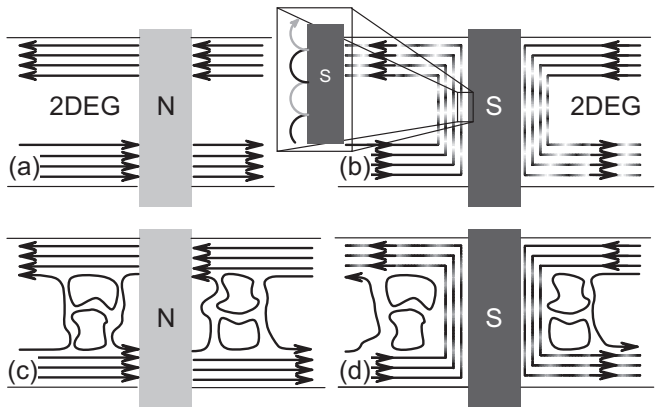}
\caption{Edge channels in a 2DEG hosting a normal (left) or superconducting (right) electrode. (a), (b): Integer filling factor (i.e. resistance minimum). (c), (d): In between integer filling factors. With a normal electrode, only the innermost channel is backscattered due to impurities in the 2DEG. In the superconducting case, edge channels hitting the electrode are Andreev reflected (see inset) and contain electrons and holes (gray). The amount of current which is backscattered depends on the hole probability, which oscillates in a magnetic field.}
\label{Fig:EdgeChannels}
\end{figure}
Fig.~\ref{Fig:EdgeChannels} shows a unit cell of the periodic arrangement of metallic stripes in a
2DEG. The edge-channel picture for normal and superconducting stripes is illustrated,
both for a full and a half-filled Landau level. A normal-conducting metal stripe acts as an ideal
contact for electrons propagating in edge channels once the stripe length greatly exceeds the cyclotron
radius. This is indeed the case for our experiment. When the stripe is
superconducting, the gap for quasiparticle excitations prevents the absorption of a
single electron and leads to Andreev reflection 
instead. As both electrons and Andreev-reflected holes
are forced on cyclotron orbits having the same chirality (see Fig.~\ref{Fig:EdgeChannels},
inset), an edge channel is formed along a superconducting contact, consisting of a coherent
superposition of electron and hole states~\cite{Bib:Zuelicke,Bib:Asano,Bib:HochB} which is stable along
its entire length. The charge current in such an Andreev edge channel is proportional to the difference
between the moduli of electron and hole amplitudes in that superposition~\cite{Bib:Zuelicke}. For an
ideal 2DEG-S-interface, Andreev reflection 
is perfect and the
Andreev edge channel is composed of electrons and holes in equal proportions. In that case, no net
current is flowing along the superconductor edge and the mesa edges remain decoupled, i.e. the
behaviour of a normal quantum Hall sample is recovered. However, when an interface
barrier and/or a Fermi-velocity mismatch leads to a finite amount of normal reflection,
interference between normal and Andreev-reflected quasiparticles 
results in $1/B$-periodic oscillations of the electron and hole amplitudes~\cite{Bib:Zuelicke,Bib:Asano}. 
For nonequal electron and hole amplitudes, a finite current is flowing parallel to the superconducting stripe,
which is fed into the normal edge channel at the opposite mesa edge and gives rise to backscattering 
between the normal edge channels.
Formally, these amplitudes can be calculated by matching
appropriate solutions of the Bogoliubov-de Gennes equations~\cite{Bib:BdG} at the interface~\cite
{Bib:Zuelicke}. A parameter $w$ was defined in Ref.~\onlinecite{Bib:Zuelicke} for characterizing the
interface barrier, which corresponds to $2Z$ in the BTK-model~\cite{Bib:BTK}. The hole probability --
and therefore the strength of the backscattering -- oscillates strongly when $w>0$, with the same
periodicity as SdH-oscillations. This is what we would expect to occur in our samples, because even
though the interface is highly transparent, there is still a residual barrier.

How can the formation of Andreev edge channels at an imperfect interface explain the enhanced
magneto-oscillations observed in our measurements? In a quantum Hall sample with normal electrodes,
the amplitude of the SdH-oscillations is determined by backscattering of the innermost edge channel
only, i.e. the one which is formed by the bulk Landau level closest to the Fermi energy. Therefore the
conductivity of the sample can only oscillate by one conductance quantum. If the electrodes are
superconducting, {\em all} edge channels are subject to Andreev reflection 
when they hit the electrode.
The oscillation amplitude is therefore not limited to one conductance quantum. For example, for $w=1$
and $\nu=18$ (which would correspond to the critical field of the Nb stripes at 1.6~K), an amplitude of
about six conductance quanta was obtained in Ref.~\onlinecite{Bib:Zuelicke}. The presence of the
screening current in the superconductor~\cite{Bib:screen} and disorder~\cite{Bib:dirt}
does not change this behavior qualitatively for edge channels whose corresponding cyclotron radius is
larger than the penetration depth but smaller than the mean free path. These conditions are satisfied
for the range of filling factors where the enhanced magnetooscillations are observed in our sample.

In the model treated in Ref.~\onlinecite{Bib:Zuelicke}, the edge channels moving along the mesa edge and hitting the 2DEG-S interface consist of electrons only, since they originate from a normal conducting electrode. Our samples incorporate many S-2DEG-S-contacts in series. For short stripe separation, the edge channels impinging on the 2DEG-S-boundary thus contain both electrons and holes. The situation of Ref.~\onlinecite{Bib:Zuelicke} is therefore not realised ideally, backscattering is less effective and the oscillations are not as pronounced. With increasing distance between the Nb stripes, more and more holes in the edge channel along the mesa edge recombine with the electrons, resulting in an edge channel containing only electrons as treated in Ref.~\onlinecite{Bib:Zuelicke}. This explains qualitatively why the oscillation amplitude increases with increasing stripe separation.

% Mechanism of the higher resistance at high fields 
Now we turn to the non-oscillatory part of the magnetoresistance. The magnetic field position of the crossover point (arrows in Fig.~\ref{Fig:TCurves}) for 16 samples was
well described by the condition $R_\mathrm{c} = 0.8\ b$, where $b=a-d$ is the distance between the stripes and $2R_\mathrm{c}$  is the cyclotron diameter in the 2DEG. 
No satisfactory dependence on either the lattice period $a$ or the stripe width $d$ was found. Additionally, the 
slope of the magnetoresistance trace (left inset in Fig.~\ref{Fig:TCurves}) changes at $2R_\mathrm{c} = b$ (i.e. one cyclotron orbit fits in between two stripes), which marks the transition to the regime of edge channel transport. 
The latter is found both above and below \Tc, hence this feature is unrelated to superconductivity.
Note that this ballistic picture is justified as the mean
free path is much larger than the perimeter of a cyclotron orbit at 
that field. Since both the 
crossing point and the change in slope are linked to the distance between the stripes, we conclude that both features are caused by the transition to the edge channel regime.

In the given geometry, we measure a series connection of many two-point resistances (metal-2DEG), shunted to an unknown fraction by the semiconductor underneath the niobium stripes. Although it is therefore difficult to make quantitative statements about the resistance we can explain qualitatively why the high-field resistance in the superconducting case is higher than in the normal case.

The Hall voltage is shunted by the metallic stripes connecting both sides of the Hall bar. This leads to a quadratic magnetoresistance, which is less pronounced in the normal state when the niobium stripes have a finite resistance. This description is valid at low fields. At high fields however, the magnetoresistance appears to be linear in $B$, as one would expect for the two-point resistance in the edge channel regime.
The two-point resistance is determined by the number of edge channels (which decreases as $B$ increases) and their conductivity, which is constant ($2e^2/h$) in a conventional quantum Hall sample. As we have seen above, the edge channels emitted by the superconducting electrodes consist of electrons and holes travelling in the same direction. Therefore, the conductivity of an edge channel is reduced compared to the normal case, which also leads to an increased resistance. 

% Summary

To summarize, we have examined arrays of Nb-filled grooves in an InAs-AlGaSb 
heterostructure at high magnetic fields using magnetotransport 
measurements at various temperatures. We observe strong $1/B$-periodic
resistance oscillations when the Nb stripes get superconducting. They are due to edge 
channels containing both electrons and holes.
We also find that above a geometry-dependent magnetic field, the overall sample resistance is higher 
in the superconducting case than in the normal case. This finding is consistent with the 
picture of edge channels containing Andreev reflected holes.
Our experiments therefore explore the impact of Andreev reflection on 
transport in the quantum Hall regime.

% Acknowledgments

The authors wish to thank J. Keller, K. Richter, and C. Strunk 
for stimulating discussions.
Financial support by the Deutsche Forschungsgemeinschaft is gratefully 
acknowledged.


\begin{thebibliography}
\bibitem{Note:Normalize} This number is a worst-case estimate assuming that the increase 
in the non-oscillatory part is only due to shorting of the Hall voltage.
\end{thebibliography}
\end{document}